\documentclass[prl,reprint,superscriptaddress,floatfix,longbibliography]{revtex4-1}

\usepackage{graphicx}
\usepackage[colorlinks=true,citecolor=blue]{hyperref}
\usepackage{amsmath}
\usepackage{amssymb}
\usepackage{soul}
\usepackage{nicefrac}
\usepackage[autostyle]{csquotes}
\usepackage{dsfont}

\usepackage{bm}

\DeclareGraphicsExtensions{.png,.jpg,.eps}
\usepackage{xcolor}

\begin{document}

\author{Viliam Va\v{n}o}
\thanks{These two authors contributed equally}
\affiliation{Department of Applied Physics, Aalto University, FI-00076 Aalto, Finland}

\author{Somesh Chandra Ganguli}
\thanks{These two authors contributed equally}
\affiliation{Department of Applied Physics, Aalto University, FI-00076 Aalto, Finland}

\author{Mohammad Amini}
\affiliation{Department of Applied Physics, Aalto University, FI-00076 Aalto, Finland}

\author{Linghao Yan}
\affiliation{Department of Applied Physics, Aalto University, FI-00076 Aalto, Finland}
\affiliation{Institute of Functional Nano and Soft Materials (FUNSOM), Jiangsu Key Laboratory for Carbon-Based Functional Materials and Devices, Soochow University, Suzhou, 215123, Jiangsu, PR China}

\author{Maryam Khosravian}
\affiliation{Department of Applied Physics, Aalto University, FI-00076 Aalto, Finland}

\author{Guangze Chen}
\affiliation{Department of Applied Physics, Aalto University, FI-00076 Aalto, Finland}

\author{Shawulienu Kezilebieke}
\affiliation{Department of Physics, Department of Chemistry and Nanoscience Center, 
University of Jyväskyl\"a, FI-40014 University of Jyväskyl\"a, Finland}

\author{Jose L. Lado}
\email{Corresponding authors. Email: jose.lado@aalto.fi, peter.liljeroth@aalto.fi}
\affiliation{Department of Applied Physics, Aalto University, FI-00076 Aalto, Finland}

\author{Peter Liljeroth}
\email{Corresponding authors. Email: jose.lado@aalto.fi, peter.liljeroth@aalto.fi}
\affiliation{Department of Applied Physics, Aalto University, FI-00076 Aalto, Finland}

\title{Evidence of nodal superconductivity in monolayer 1H-TaS$_2$ with hidden order fluctuations} 

\begin{abstract}

Unconventional superconductors represent one of the fundamental directions in modern quantum materials research. In particular, nodal superconductors are known to appear naturally in strongly correlated systems, including cuprate superconductors and heavy-fermion systems. Van der Waals materials hosting superconducting states are well known, yet nodal monolayer van der Waals superconductors have remained elusive. Here, using low-tempera\-ture scanning tunneling microscopy (STM) and spectroscopy (STS) experiments, we show that pristine monolayer 1H-TaS$_2$ realizes a nodal superconducting state. By including non-magnetic disorder, we drive the nodal superconducting state to a conventional gapped s-wave state. Furthermore, we observe the emergence of many-body excitations close to the gap edge, signalling a potential unconventional pairing mechanism. Our results demonstrate the emergence of nodal superconductivity in a van der Waals monolayer, providing a building block for van der Waals heterostructures exploiting unconventional superconducting states.

\end{abstract}


\date{\today}

\maketitle

\renewcommand\thesection{\arabic{section}}


Unconventional superconductivity is associated with the existence of competing electronic interactions and it represents one of the paradigmatic correlated states of matter. A variety of compounds have been discovered in this broad family in the last decades, including high-temperature superconductors, heavy-fermion superconductors and (artificial) topological superconductors \cite{RevModPhys.83.1589,Allan2013,Zhou2013,Wirth2016,Stewart2017,Jiao2020,RevModPhys.66.763,RevModPhys.74.235,UPt32014,UPt32020,NadjPerge2014,TSC2020}. While a wide range of unconventional superconducting states have been found in bulk and complex compounds and van der Waals materials hosting superconductivity are well-known \cite{Gamble1971,PhysRevLett.28.299,Xi2015,Navarro2016,delaBarrera2018,Cao2018,BoixConstant2021}, intrinsic unconventional superconductivity in a monolayer van der Waals material has not yet been discovered \cite{Ribak2020,Nayak2021,Oh2021,kim2021spectroscopic}. One of the best known monolayer superconductors, NbSe$_2$, is a conventional s-wave superconductor in the monolayer limit \cite{Ugeda2015,PhysRevB.99.161119,Khestanova2018}. Despite its conventional nature, recent results point towards the proximity to potential correlated ground states \cite{PhysRevB.89.224512,2021arXiv210104050W,Ganguli2022,PhysRevX.10.041003}, suggesting that other related  materials would be outstanding candidates for monolayer unconventional superconductors.

Strongly correlated superconductors are prone to show unconventional order parameters \cite{RevModPhys.63.239}. While phonon-driven
superconductors present s-wave superconducting symmetry, coexistence of strong electronic correlations
can energetically favour higher angular momentum states. In particular, high-Tc cuprate superconductors, in which the
pairing mechanism is driven by magnetic excitations, present nodal d$_{x^2-y^2}$-wave superconducting order parameter,
compatible with the underlying $C_4$ symmetry of the electronic structure. Interestingly, correlated superconductors
in triangular lattices with $C_3$ rotational symmetry would be prone to other unconventional superconducting order parameters, as observed in heavy-fermion superconductors \cite{RevModPhys.74.235,UPt32014,UPt32020}.
While the microscopic mechanism leading to these states represents an open research question, correlated superconductors
often show two distinctive signatures simultaneously: an unconventional order parameter with nodal superconducting states, and the appearance of low-energy many-body fluctuations coexisting with the superconducting state \cite{PhysRevB.67.224502}. In addition, unconventional superconducting states are well-known to be highly sensitive to non-magnetic disorder, a feature commonly used as a probe of unconventional superconductivity \cite{PhysRevLett.80.161,RevModPhys.78.373}. Here we show that clean monolayer 1H-TaS$_2$ simultaneously presents an unconventional nodal superconducting state and low energy many-body fluctuations, hallmarks of a correlated superconducting state. Furthermore, non-magnetic disorder drives the system to a conventional superconducting state with s-wave pairing symmetry. Such realization of unconventional superconductivity in a simple monolayer van der Waals material allows its detailed study under external stimuli (e.g., gating). In addition, this adds a crucial building block for designer van der Waals heterostructures exploiting unconventional superconductivity.
\begin{figure*}[t!]
    \centering
    \includegraphics[width=.99\textwidth]{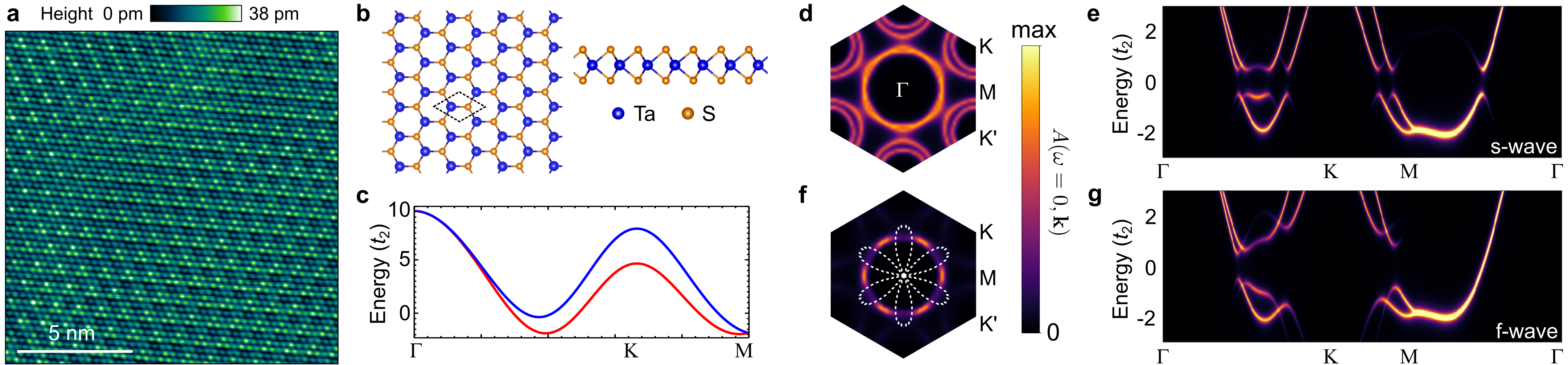}
	\caption{Superconducting pairing symmetry in 1H-TaS$\mathbf{_2}$. a,b) STM image (a, set point parameters $V = 10$ mV and $I$ = $1$ nA) and schematic structure (b) of 1H-TaS$_2$. c,d) Calculated band structure (c), and the Fermi surface (d) of monolayer 1H-TaS$_2$. Blue and red in (c,d) correspond to up and down channels, with the splitting stemming from Ising spin-orbit coupling. e) Including s-wave pairing opens a superconducting gap uniformly across the Fermi surface in the calculated electron spectral function. f,g) Calculated Fermi surface (f) and the electron spectral function (g) with nodal pairing. The dotted line in panel (f) shows the symmetry of the nodal superconducting pairing.}
    \label{fig:schem}
\end{figure*}

\section{Results and discussion}
\textbf{Superconductivity in a pristine 1H-TaS$_2$ monolayer.} We grow 1H-TaS$_2$ on highly oriented pyrolytic graphite (HOPG) using molecular beam epitaxy (MBE) (for growth details see the Methods section) and investigate its electronic structure using low-temperature scanning tunneling microscopy (STM) and spectroscopy (STS) experiments. The MBE growth results in clean samples with a low defect density and the 1H-TaS$_2$ monolayer exhibits the well-studied $3\times3$ charge density wave (STM image shown in Figure \ref{fig:schem}a and schematic of 1T-TaS$_2$ atomic structure in Figure \ref{fig:schem}b) \cite{PhysRevB.94.081404,Lin2018,Hall2019}. Larger bias range spectroscopy shows states arising from metal d-band (Supporting Information (SI) Figure S1) similarly to the case of 1H-NbSe$_2$ \cite{Ugeda2015}.

While superconductivity in thin films of 2H-TaS$_2$ down to monolayer thickness has been observed \cite{Navarro2016,delaBarrera2018,PhysRevB.98.035203,Peng2018,Bekaert2020}, the gap symmetry of monolayer 1H-TaS$_2$ has not been established to date \cite{Guillamon2011,PhysRevB.89.224512}. The scenario for realizing unconventional superconductivity in 1H-TaS$_2$ can be illustrated by tight-binding calculations (details in the SI Section S2). The broken inversion symmetry is reflected in the band structure of this material (Figure \ref{fig:schem}c) in the form of Ising spin-orbit coupling, leading to enormous band splitting around K and K' points (Figure \ref{fig:schem}c,d). Introducing s-wave pairing potential gaps the Fermi surface of 1H-TaS$_2$ uniformly in reciprocal space (Figure \ref{fig:schem}e) leading to conventional superconductivity. It is also possible to construct unconventional superconducting states in 1H-TaS$_2$. The symmetry of the system is consistent with nodal f-wave pairing \cite{PhysRevB.89.224512}, which has a gap amplitude in reciprocal space analogous to the $f_{y(y^2-3x^2)}$ orbital (as discussed in Sec.~S2B-D in the SI), as illustrated by white dashed line in Figure \ref{fig:schem}f. F-wave pairing potential gaps the Fermi surface along $\Gamma$-K/K', but not along $\Gamma$-M direction (Figure \ref{fig:schem}f,g). In contrast to s-wave symmetry giving a fully gapped spectrum, this leads to gapless, nodal superconducting state. The possible presence of doping or strain is not expected to change this conclusion (see SI Section S2E for details).

F-wave pairing is not the only possible nodal order parameter. For example, s$^{\pm}$ order parameter would also be consistent with the nodal spectra observed (see below for experimental results). However, we consider f-wave order parameter as a more likely scenario as signatures of it have been observed in monolayer dichalcogenides \cite{2021arXiv210104050W,PhysRevB.89.224512}. In contrast, no signatures of s$^{\pm}$ order have been observed in this type of materials. Comparison with an s$^{\pm}$ is presented in the SI Section S2F. Finally, in the presence of spin-orbit coupling, singlet and triplet orders can coexist. We analyse this scenario in SI Section S2G, and conclude that it should not lift the presence of the superconducting nodes in the electronic structure.

\begin{figure*}[t!]
    \centering
    \includegraphics[width=.99\textwidth]{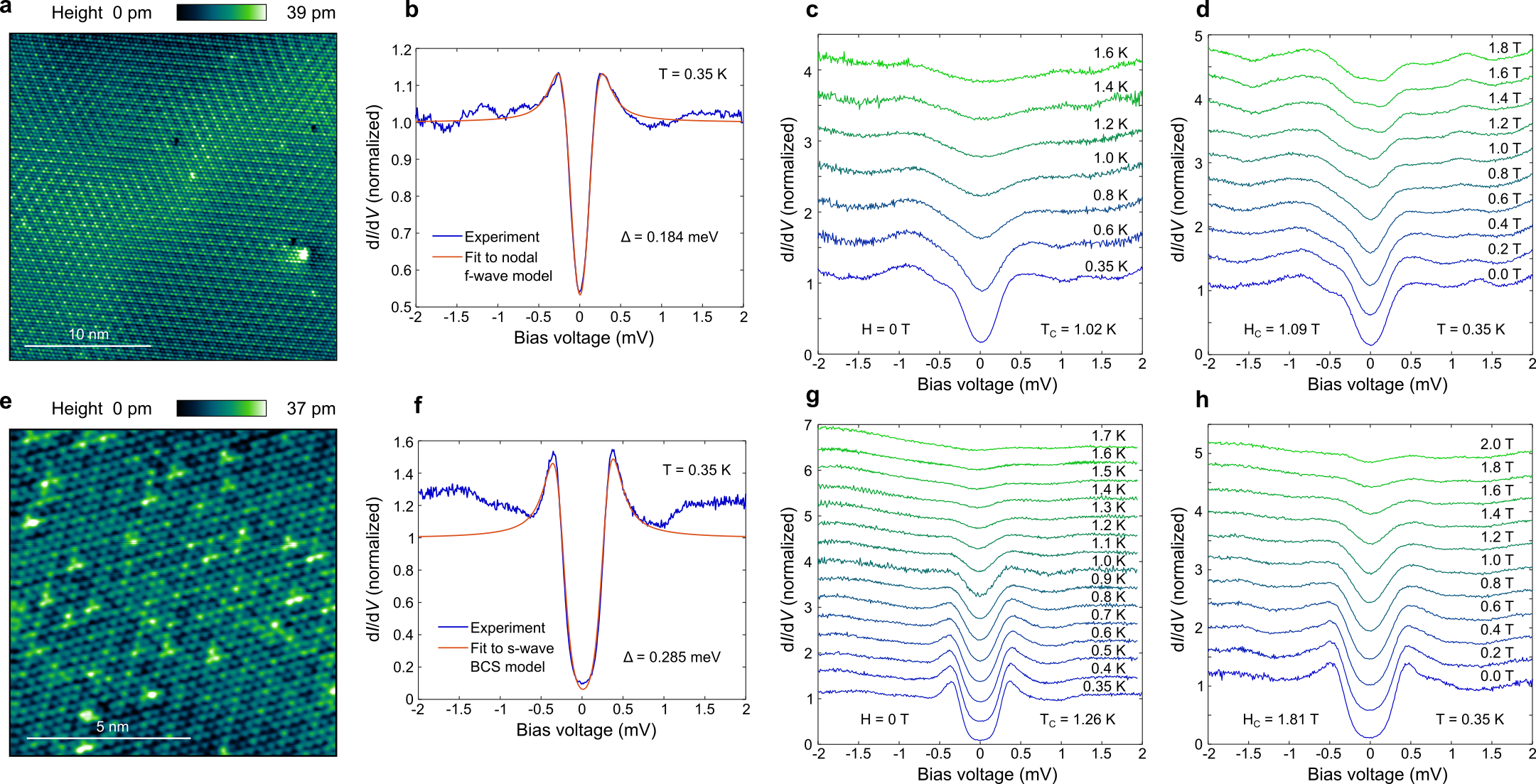}
	\caption{Superconductivity in 1H-TaS$\mathbf{_2}$. a-d) STM/STS results on a clean 1H-TaS${_2}$ sample. a) STM image of 1H-TaS$_2$ on HOPG substrate, obtained at $V = 10$ mV and $I$ = $1$ nA. b) d$I$/d$V$ spectrum of the clean sample (blue), averaged over 12 spectra measured on different islands. Orange curve represents a fit using nodal order parameter. c,d) Temperature (c) and magnetic field (d) dependence of a superconducting gap of the clean sample, spectra are vertically offset for clarity. e-h) STM/STS results on a disordered 1H-TaS${_2}$ sample. e) STM image of the disordered 1H-TaS$_2$ on HOPG substrate, obtained at $V = 10$ mV and $I$ = $1$ nA. f) d$I$/d$V$ spectrum of the disordered sample (blue), orange curve represents a fit using s-wave BCS model. g,h) Temperature (g) and magnetic field (h)  dependence of a superconducting gap of the disordered sample, spectra are vertically offset for clarity. d$I$/d$V$ spectra used in this figure were performed with a bias modulation of $V_\mathrm{mod} = 20$ $\mu$V. d$I$/d$V$ spectra in panels (c,d,f,g,h) are an average over 4 $\times$ 4 grid with size 5 nm.}
    \label{fig:SC}
\end{figure*}

As illustrated in Figure \ref{fig:SC}a,b, high-energy resolution d$I$/d$V$ spectrum on a clean 1H-TaS$_2$ sample reveals a V-shaped gap at the Fermi level. We confirm that this gap is of superconducting origin by measuring its temperature and magnetic field dependence (Figure \ref{fig:SC}c,d). The spectral gap vanishes with increasing temperature and out of plane magnetic field, which is consistent with the superconducting nature of the gap. We did not observe a vortex lattice on our sample (also not observed on related 1H-NbSe$_2$ monolayer superconductor \cite{Zhao2019,Ganguli2022}).

It is immediately clear that the gap shape is not consistent with s-wave BCS model. While this tunneling spectrum could also be fitted using s-wave Dynes model that includes pair-breaking processes through the Dynes parameter $\Gamma$ \cite{PhysRevLett.41.1509}, we argue that the physics of this model is inconsistent with our experiments. For an s-wave superconductor, pair-breaking would require magnetic impurities to account for physics behind the $\Gamma$ in the Dynes model. This is contradictory to having an ultraclean sample \cite{Herman2016}. Secondly, $\Gamma$ and the superconducting gap $\Delta$ should be anticorrelated within the Dynes model and this is in contrast with our experiments (see SI Figure S8). Furthermore, the disordered sample (see below) was found to fit with $\Gamma = 0$ meV (i.e. pure s-wave BCS model). It would be physically inconsistent that increasing disorder removes the pair-breaking processes coming from impurities. In stark contrast, the data is fully accounted for by a model with nodal superconductivity without introducing the somewhat artificial Dynes parameter (see Figure \ref{fig:SC}b). Details about the fitting procedure as well as fits using other models for tunneling spectroscopy of superconductors can be found in the SI Section S3. Finally, we note that there are some differences in the shape of the coherence peaks in Figs.~\ref{fig:SC}b and c as these data are from different experimental runs, which can be due to changes in the microscopic structure of the STM tip. However, this does not affect our conclusion on the shape of the gap and whether it can be fit with a nodal or s-wave BCS model.

\textbf{Superconductivity in a disordered 1H-TaS$_2$ monolayer.} Conventional superconductors are robust against weak non-magnetic impurities. This feature, known as
Anderson's theorem, also implies that under weak disorder, conventional s-wave superconducting gaps are
not substantially affected \cite{Anderson1959}.
The situation is different for unconventional superconductors, in which
non-magnetic impurities are known to drastically impact the spectral properties. In particular,
non-magnetic disorder alone can create in-gap states in an unconventional superconducting gap. This behaviour
can be rationalized from the fact that non-magnetic impurities create scattering in reciprocal space, and due
to the sign-changing nature of unconventional superconducting orders, disorder induces
an average zero superconducting order in reciprocal space.

\begin{figure*}[t!]
    \centering
    \includegraphics[width=.99\textwidth]{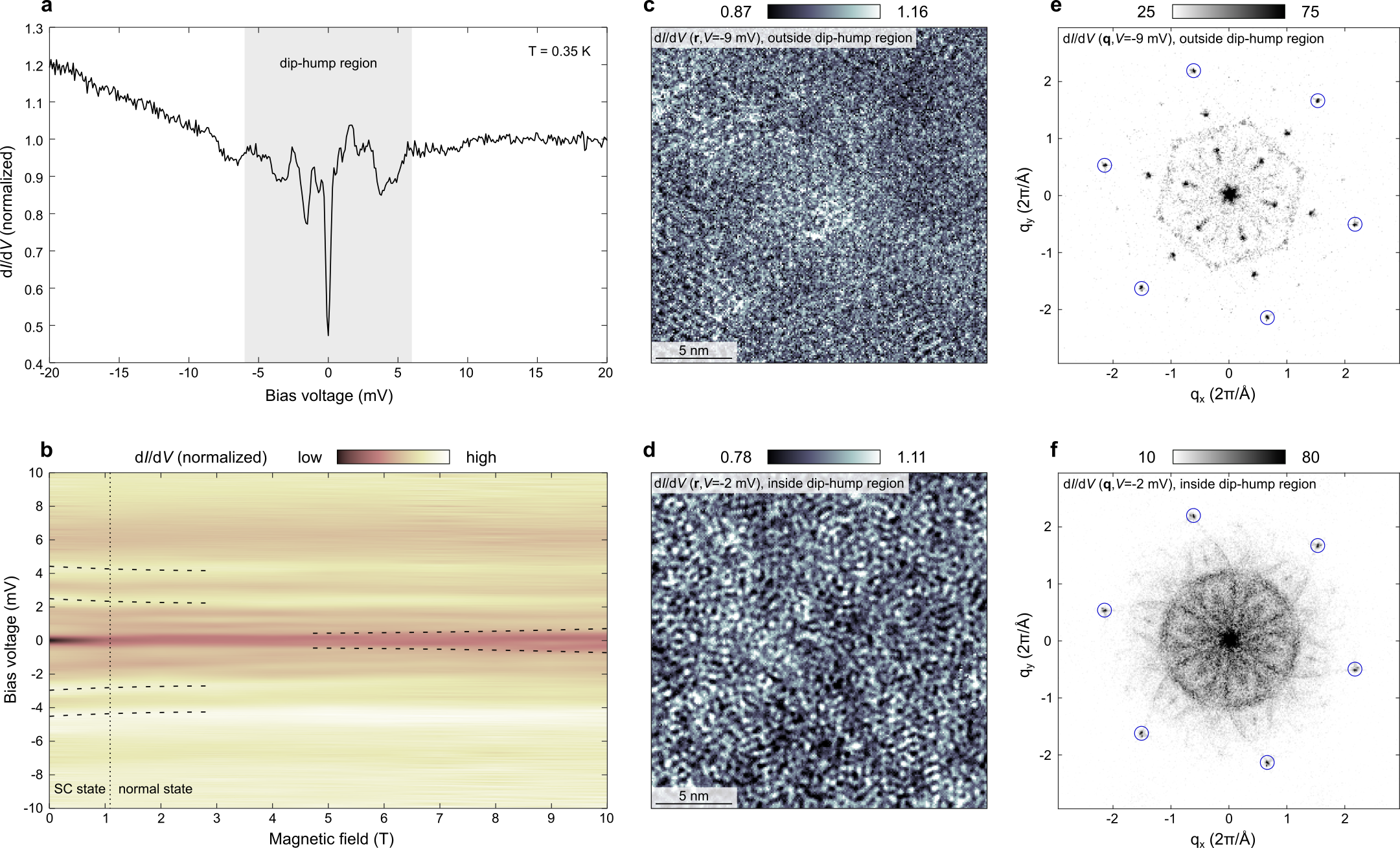}
	\caption{Dip-hump features around the Fermi level of 1H-TaS$\mathbf{_2}$. a) d$I$/d$V$ spectrum of 1H-TaS${_2}$ displaying fluctuations in $\sim\pm6$ mV window (shaded region) and a smooth evolution of the LDOS outside this region ($V_\mathrm{mod} = 200$ $\mu$V). b) Magnetic field dependence of the d$I$/d$V$ spectroscopy ($V_\mathrm{mod} = 100$ $\mu$V). Dashed lines highlight the main features that vary with magnetic field, the dotted line marks the critical field of superconductivity. The spectrum is an average over 4 $\times$ 4 grid with size 5 nm. c,d) d$I$/d$V$ maps of the same area at $V = -9$ mV (c) and at $V = -2$ mV (d) ($V_\mathrm{mod} = 400$ $\mu$V). e,f) Fourier transforms of d$I$/d$V$ maps at $V = -9$ mV (e) and at $V = -2$ mV (f). The Fourier transforms have been three fold symmetrized, respecting the C$_3$ symmetry of the lattice. Results presented in this figure were measured on a clean sample. All the results presented in this figure were measured at $T = 0.35$ K.}
    \label{fig:wiggles}
\end{figure*}

Introducing impurities on our sample makes the $3 \times 3$ CDW disordered, which can be seen on Figure \ref{fig:SC}e (more details in the SI Section S4). Interestingly, while we observe that clean 1H-TaS$_2$ shows a nodal-like superconducting order, the situation with disorder is drastically different. In particular, the superconducting gap recovers the fully gapped form of a conventional superconductor, that is well-fit by an s-wave BCS model (Figure \ref{fig:SC}f). This phenomenology can be rationalized from the impurity physics associated to unconventional superconductors (see SI Section S2H for details). In the pure limit, the system shows an unconventional nodal superconducting order. When impurities are introduced, the reciprocal space scattering quickly quenches the unconventional superconducting state. In this disordered limit, a competing s-wave superconducting order takes over, as it is not affected by the presence of non-magnetic disorder (see SI Section S2H for details). The drastically different spectral shapes support the transition from a gapless superconducting order in the clean limit to a conventional gapped s-wave order in the disordered limit. 

It can also be observed that the s-wave gap is larger than the clean nodal one, with a concomitant increase in $T_c$ and $H_c$ as well. The fact that we have two competing order parameters implies that their associated critical temperatures are similar, which has been observed in other correlated materials \cite{Grinenko2020}. The enhancement of the gap can be rationalized in terms of impurity-induced electronic-structure reconstruction \cite{PhysRevLett.98.027001} and dominance of a disorder-induced inhomogeneous conventional channel over the pristine unconventional channel \cite{PhysRevLett.81.3765,Kaneyasu2010}. This phenomenology arises from the resilience of conventional s-wave to disorder and the vulnerability of the unconventional nodal order, leading to the prevalence of s-wave superconductivity in disordered samples. Once again, as expected, increasing the temperature or magnetic field suppresses the spectral gap with conventional s-wave character (Figure \ref{fig:SC}f,g). Details on the determination of the $T_c$ and $H_c$ can be found in SI Section S5, spatial variation of the superconducting gap in Section S6.

\textbf{Many-body excitations around the superconducting gap.} In addition to the superconducting gap, we observe low energy excitations in the d$I$/d$V$ spectra as shown in Figure \ref{fig:wiggles}a. While these features are associated to electron-boson coupling, their energies are not consistent with the known phonon energies and electron-phonon couplings in 1H-TaS$_2$ \cite{Hinsche2017,Wijayaratne2017}. They are likely to arise from many-body excitations associated with the superconducting state. Similar dip-hump features were found in a closely related 1H-NbSe$_2$ system \cite{2021arXiv210104050W}. However, in contrast to the results on 1H-NbSe$_2$, the excitations in 1H-TaS$_2$ do not have a constant energy spacing, they are not fully symmetric around the superconducting gap and persist at higher magnetic fields in the normal state (Figure \ref{fig:wiggles}b, more data in the SI Section S7). Here, upon applying magnetic field, the dip-hump features shift to lower energies at low magnetic fields and remain at a constant energy at higher magnetic fields (above the critical field of $\sim 1$ T \cite{delaBarrera2018}, see Figure \ref{fig:SC}d). 

Similar excitations in high-Tc superconductors are believed to stem from an antiferromagnetic fluctuating order, responsible for the pairing mechanism in those compounds \cite{PhysRevB.67.224502}. The excitation peaks in our data can be analogously associated to an underlying fluctuating order, potentially responsible for the nodal superconducting state (see discussion in SI Section S2I).  
These excitations have a clear fluctuation length scale in both real and reciprocal space. Outside the dip-hump region, highlighted in  Figure \ref{fig:wiggles}a, the tunneling conductance has low spatial fluctuations (Figure \ref{fig:wiggles}c), while inside the region the spatial fluctuations are much stronger (Figure \ref{fig:wiggles}d, can also be seen on Figure S16). In the reciprocal space, there are several features both inside and outside the dip-hump region. Outside (Figure \ref{fig:wiggles}e), there are peaks that correspond to lattice (blue circles), to the $3 \times 3$ CDW with their satellites, and there is a hexagonal shape probably corresponding to quasiparticle interference of the 1H-TaS$_2$ d-band. Inside (Figure \ref{fig:wiggles}f), the lattice peaks and CDW peaks are suppressed, while there appears a strong flower-shaped pattern. The reciprocal space features are very different for outside and inside the dip-hump region and this behaviour is present in all measurements (for full datasets, see Supporting Movies 1-3). While the nature of the many-body
excitations cannot be fully determined at the current stage, their concomitant appearance with the nodal superconducting gap suggests that they can potentially
be assiciated with the pairing channel 
superconducting state. Further work should be carried out to establish its microscopic nature
and exact connection with nodal state in TaS$_2$.

Another feature associated with the unconventional superconducting order is a pseudogap phase appearing in the vicinity of a superconducting dome \cite{Timusk1999}. In our experiments, when superconductivity is quenched by an external magnetic field, a pseudogap appears at the Fermi level. This can be seen at higher magnetic fields in Figure \ref{fig:wiggles}b with the pseudogap getting wider with increasing field. 

\section{Conclusions}
We have provided evidence for unconventional nodal superconductivity in ultra-clean monolayer 1H-TaS$_2$ by means of STM and STS measurements. In particular, we have demonstrated three genuine signatures of an unconventional superconducting state: a nodal V-shaped superconducting gap incompatible with a BCS gap, many-body excitations around the Fermi level potentially associated with the unconventional superconducting order, and the emergence of a pseudogap in the normal state with an applied magnetic field. In addition, introduction of disorder in 1H-TaS$_2$ gives rise to a conventional s-wave superconducting state, further supporting the unconventional nature of superconductivity in ultra-clean samples. Beyond the fundamental interest in unconventional monolayer superconductors, the vdW nature of this material implies that 1H-TaS$_2$ can be used as a building block to incorporate this new electronic order into artificial van der Waals heterostructures. Ultimately, the discovery of a monolayer unconventional superconductor opens up potential new strategies to realize a whole new family of exotic states relying on nodal superconductivity by stacking and twisting various vdW materials.

\section*{Experimental Section}
\emph{Sample preparation.} 
TaS$_2$ was grown by molecular beam epitaxy (MBE) on highly oriented pyrolytic graphite (HOPG) under ultra-high vacuum conditions (UHV, base pressure $\sim1\times10^{-10}$ mbar). HOPG crystal was cleaved and subsequently out-gassed at $\sim800^\circ$C. High-purity Ta and S were evaporated from an electron-beam evaporator and a Knudsen cell \cite{Hall2018}, respectively. Prior to growth, the flux of Ta was calibrated on a Au(111) at $\sim1$ monolayer per hour. The ratio of 1T to 1H-TaS$_2$ can be controlled via substrate temperature and the overall coverage \cite{Lin2018}. Before the growth, HOPG substrate temperature was stabilised at $\sim500^\circ$C, at which only the 1H phase grows. The sample was grown in a S pressure of $\sim5\times10^{-8}$ mbar and the growth duration was 25 minutes. 

\emph{STM measurements.} After the sample preparation, it was inserted into the low-temperature STM (Unisoku USM-1300) housed in the same UHV system and subsequent experiments were performed at $T = 350$ mK. STM images were taken in the constant-current mode. d$I$/d$V$ spectra were recorded by standard lock-in detection while sweeping the sample bias in an open feedback loop configuration, with a peak-to-peak bias modulation specified for each measurement and at a frequency of 911 Hz. d$I$/d$V$ spectra were recorded on extended monolayers to eliminate possible island size dependence from the measurements. The direction of the applied magnetic field is out of plane.

\section*{Acknowledgements}
This research made use of the Aalto Nanomicroscopy Center (Aalto NMC) facilities and was supported by the European Research Council (ERC-2017-AdG no.~788185 ``Artificial Designer Materials'') and Academy of Finland (Academy professor funding nos.~318995 and 320555, Academy research fellow nos.~331342, 336243 and no.~338478 and 346654). L.Y. acknowledges support from the Jiangsu Specially-Appointed Professors Program, Suzhou Key Laboratory of Surface and Interface Intelligent Matter (Grant SZS2022011), Suzhou Key Laboratory of Functional Nano \& Soft Materials, Collaborative Innovation Center of Suzhou Nano Science \& Technology, and the 111 Project. We acknowledge the computational resources provided by the Aalto Science-IT project.

\bibliography{biblio} 

\end{document}